\documentclass[twocolumn,a4paper,aps,prl,showpacs]{revtex4}
\usepackage[windows]{umlaute}
\usepackage[dvips]{epsfig}

\newcommand{\ptwo}{p$(2\times 2)$}
\newcommand{\one}{$(1\times 1)$}
\begin{document}

\title{Role of elastic scattering in electron dynamics at ordered alkali
overlayers on Cu(111)}

\author{
C. Corriol,$^1$
V. M. Silkin,$^1$
D. S\'anchez-Portal,$^{1,2}$
A. Arnau,$^{1,2,3}$
E. V. Chulkov,$^{1,2,3}$
P. M. Echenique$^{1,2,3}$}

\affiliation{
$^1$Donostia International Physics Center (DIPC),
$^2$Centro Mixto C.S.I.C.-UPV/EHU, and 
$^3$Departamento de F\'{\i}sica de Materiales,
Facultad de Qu\'{\i}mica, Apartado 1072, San Sebastian 20080,
Spain}

\author{T. von Hofe}
\author{J. Kliewer}
\altaffiliation{Present address:
Infineon Technologies AG, D-81609 München, Germany}
\author{J. Kröger}
\author{R. Berndt}
\affiliation{
Institut für Experimentelle und Angewandte Physik,
Christian-Albrechts-Universität zu Kiel,
D-24098 Kiel,
Germany}

\begin{abstract}

Scanning tunneling spectroscopy of \ptwo\ Cs and Na ordered overlayers
on Cu(111) reveals similar line widths of quasi two-dimensional
quantum well states despite largely  different binding energies.
Detailed calculations show  that 50 \% of the line widths are
due to electron-phonon scattering while inelastic electron-electron
scattering is negligible.  A frequently ignored mechanism for ordered
structures, i.e., enhanced elastic scattering due to
Brillouin zone back folding, contributes the remaining width.
\end{abstract}


\maketitle

In many electron systems ubiquitous interactions between
quasiparticles such as electron-electron, electron-phonon or
electron-magnon scattering are present and cause inelastic
scattering.  A fingerprint of this dynamics, which occurs on a
femtosecond timescale, is the width of spectroscopic lines. Often,
elastic scattering is significant too, except for well defined
surface states at special points in the surface Brillouin zone (SBZ).
Thus, surface resonances exhibit a finite width owing to a degree of
mixing with the bulk continuum.  Defect scattering can also increase the
line width of electron states at artificial atomic arrays or
at electron confining island \cite{crommie,lutz,inglesfield,hje}.
Adsorbates on a surface lead to
additional scattering channels that may change the dynamics
of electrons at surfaces and at the same time can be an important
source of elastic scattering for electrons \cite{borisov}.
An intriguing situation occurs when an extended {\em ordered}
superstructure changes the periodicity of a surface, as in the case
of submonolayer coverages of alkali atoms on (111) noble metal
surfaces. The SBZs of the substrate and of the overlayer
have different sizes. The resulting back folding of the bands
can lead to large effects in the dynamics of electrons at
the interface, like the appearance of an elastic width of quantum
well states (QWS)
when energy gaps are closed in certain regions of the SBZ.

Inverse photoemission and two-photon photoemission are widely used
techniques to probe the dynamics of hot electrons at
surfaces \cite{petek,echenique}. Recently, the different mechanisms that
contribute to the elastic and inelastic scattering of hot electrons
at Cu(001) surfaces with a low coverage of Cu adatoms have been
identified \cite{fauster} using these techniques. The conclusions are
deduced from
information taken over the whole surface area without probing the
local
electronic structure. Scanning tunneling spectroscopy (STS)
resolves the local electronic structure
of surface areas free from defects.
It also can provide detailed information about both occupied and unoccupied
states. 
The lifetime broadening of surface states can be
extracted from STS measurements by doing a line shape
analysis \cite{li,kliewer,bauer,limot} or from electron
standing wave patterns \cite{burgi,rieder,scr_05}. The progress of
these
experimental approaches clearly requires refined theoretical methods
if one aims to explain experimental data at a quantitative level.

Here, in a combined experimental and theoretical study, we
analyze the processes that contribute to the width of QWSs  
at Cs and Na overlayers on Cu(111). We achieve
quantitative agreement between low temperature STS
experiments and simulations of tunneling spectra using first
principles calculations of the electronic structure of the surface
and of the line width. The width of the QWS is shown to have
significant contributions from the elastic width that appears as a
result of the coupling to substrate states and from the inelastic
electron-phonon scattering whereas inelastic electron-electron
interaction is
of minor importance, contrary to the situation observed in studies of
the hole dynamics at the (111) surface of noble metals \cite{kliewer}.

Clean Cu(111) surfaces were prepared by standard Ar$^+$ bombardment
and annealing cycles. Cs and Na were deposited from commercial
dispensers \cite{saes} onto the sample at room temperature at rates
of $\approx 0.5\,{\rm ML}\,{\min}^{-1}$ as monitored with a quartz
microbalance. We define one monolayer (ML) as one alkali atom per one
Cu atom of the unreconstructed Cu(111) substrate. In the Cs case, a
clear \ptwo\ low-energy electron diffraction (LEED) pattern at
room temperature  \cite{CsLEED,Diehl} was used as additional
indication of
a $0.25\,{\rm ML}$ coverage. After preparation, samples were
transfered to a scanning tunneling microscope (STM) and cooled to
temperatures of $4.6-9 $ K\@.
Spectroscopy of the differential conductance (${\rm d}I/{\rm d} V$),
was performed on
homogeneous areas using a lock-in amplifier and adding a sinusoidal
modulation ($6 - 14\,{\rm mV_{pp}}$, $10\,{\rm kHz}$) to the sample
voltage.

\begin{figure}[tbp]
\resizebox{15.5cm}{!}{
\includegraphics*[60mm,2mm][320mm,205mm]{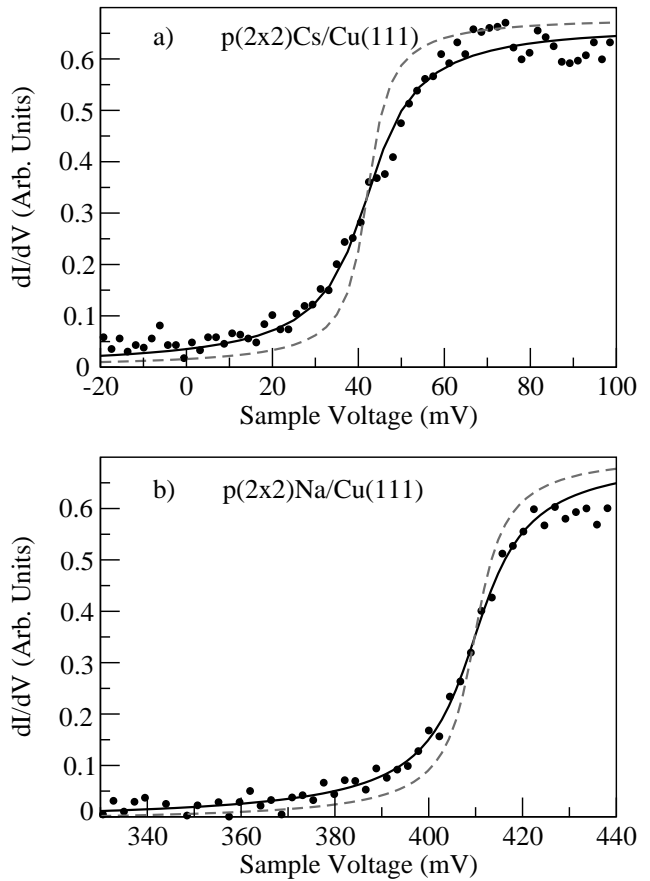}}
\caption{Comparison between measured ${\rm d}I/{\rm d} V$ vs.\ $V$
curves
(dots) for \ptwo\ structures of (a) Cs and (b) Na on Cu(111)
along with calculated results (lines).
The solid lines include
both the elastic and inelastic contributions to the energy width,
while the dashed lines include only the inelastic one.}
\label{data}
\end{figure}

Figure \ref{data} displays ${\rm d}I/{\rm d} V$-spectra (dots) of
\ptwo\ Cs and Na layers.  Sharp rises of the conductance at $E_0
\cong 40 $ meV and $E_0 \cong 410$ meV correspond to the minima of
the quantum well state bands of Cs and Na overlayers, respectively. The width of
the rise, which is related to the inverse lifetimes of the these states
\cite{li}, is similar in both cases. This observation is surprising
in view of the large difference in binding energy which, in a simple
picture, should cause widely different electron-electron ($e-e$)
scattering. However, this appears not to be the case. Our
calculations, whose results agree well with the experimental
data (solid lines in Fig.\ \ref{data}), show that inelastic $e-e$
scattering
is unimportant and that electron-phonon ($e-ph $) scattering explains
only 50 \% of the line widths. Enhanced elastic scattering due to
Brillouin zone back folding, turns out to contribute the missing 50
\% of the line width.

\par The surface electronic structure of both systems has been
calculated using the density functional code VASP~\cite{kresse}.
Projected Augmented Waves~\cite{blochl} are used to describe
the ionic cores and a plane wave basis set for the valence
electrons.
The exchange and correlation potential is treated in the generalized
gradient approximation~\cite{perdew}. We have checked the
convergency of
the calculations with respect to the thickness of the slab up to 10
layers of Cu. The $k$-point sampling of the irreducible wedge of the
surface Brillouin zone of the (2x2) superstructure is done using
14 special points for the self-consistent calculation and later on
we use up to 154 points to achieve high enough resolution of the
bands.
These slab calculations provide the structure for the recursive
calculation of the elastic width in the semi-infinite medium.
The tunneling conductance is calculated with a model that assumes a
smooth density of states of the tip around the Fermi level. Then, at
small positive sample bias, the energy dependence of the conductance
is determined by the sample density of states \cite{hormandiger}.
The details of the model are described elsewhere \cite{corriol}.
In brief, we introduce an energy width ($\Gamma_m/2$) of the states
when calculating the imaginary
part of the surface Green function, i.e., the surface local density
of states (LDOS). This width corresponds to a Lorentzian weight for
each state such that the surface LDOS is given by:

\begin{equation}
\rho(\varepsilon,\vec r)=\frac{1}{\pi}\sum_{m}|\Psi_m(\vec
r)|^2\frac{\Gamma_m/2}{(\varepsilon-\varepsilon_m)^2 +
(\Gamma_m/2)^2}
\end{equation}

The eigenvalues ($\varepsilon_m$) and eigenfunctions ($\Psi_m$) are
taken from VASP, while the energy width ($\Gamma_m/2$) is determined
as described below. The sum over {\em m} includes both a summation over
the band
index and an integration over parallel momentum. The LDOS is calculated
at 3 \AA\ from the alkali overlayer, where the electronic corrugation is
negligible, and it is dominated by the QWS band in the relevant energy
range. A $k$-point interpolation of the QWS is used to achieve the
required meV resolution of the band \cite{corriol}.

For comparing the measured  ${\rm d}I/{\rm d} V$ data for the
\ptwo\ Cs structure on Cu(111)
and the simulated spectrum we return to Fig.\ \ref{data}a. No
fitting parameter has been used apart
from a scaling factor for the calculated conductance.
The agreement between the data points and the solid curve clearly
shows
the importance of the elastic width in determining the shape of the
spectrum at the onset around 40 meV. Fig. \ref{data}b
displays a similar comparison for a \ptwo\ Na film on
Cu(111). In this case, the calculated curves have been shifted by
+18 meV
to match the experimental binding energy.  This small shift is
consistent with the expected accuracy of density functional
calculations \cite{carlsson}. 
The shape of the curve at the onset also agrees well 
with the measured data in this case.

\par
\begin{figure}[tbp]
\centering
\rotatebox{270}{
\resizebox{6.5cm}{!}{
\includegraphics*[15mm,15mm][190mm,260mm]{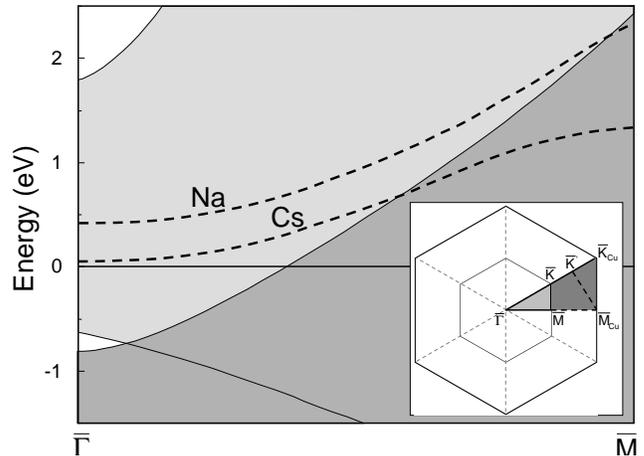}}}
\caption{Dispersion of \ptwo\ Na and Cs
quantum well states in the
$\overline{\Gamma}-\overline{{\rm M}}$ direction. Dark grey
indicates
bulk Cu states of the \one\ SBZ,
while light grey shows the closing of the gap around
$\overline{\Gamma}$ after back folding from the
$\overline{{\rm M}}-\overline{{\rm M}}_{\rm Cu}$
and $\overline{{\rm M}}_{\rm Cu}-\overline{{\rm K'}}$ symmetry
directions defined in the inset.}
\label{disp}
\end{figure}

 The adsorption of alkali adlayers in \ptwo\
superstructures leads to a modification of the projected bulk
band structure of the substrate as shown in Fig.\ \ref{disp} along
the
$\overline{\Gamma}-\overline{{\rm M}}$ direction. The inset displays
the SBZs for a
\ptwo\ overlayer and for a \one\ substrate. In the
overlayer case the energy gap near $E_F$ of clean
Cu(111) surface is closed due to the folding of bands originally
situated along the $\overline{{\rm M}}-\overline{{\rm M}}_{\rm Cu}$ and
$\overline{{\rm M}}_{\rm Cu}-\overline{{\rm K'}}$ symmetry directions. The
corresponding
extra corrugation of the Cu(111) surface due to the coupling between
the \ptwo\ adsorbate layer and underlying Cu bulk states
causes a degree of mixing between the alkali induced QWS at
$\overline{\Gamma}$ and substrate Cu states at the $\overline{{\rm
M}}_{\rm Cu}$
point of the \one\ Brillouin zone.  This serves as a
source of elastic scattering, which requires a quantitative analysis. 

\par To calculate the elastic width of QWSs
we compute the Green function of the combined system (adlayer plus
semi-infinite substrate) with high precision in the surface region.
Here, we use the SIESTA code~\cite{SIESTA2} which utilizes a basis set
of numerical atomic orbitals~\cite{basis}. The results for the
electronic
structure are almost identical to those obtained with VASP.
With this local basis set the elements of the Hamiltonian and overlap
matrices between atoms that are far apart
(beyond $\approx$15~\AA\ in our case) are strictly zero. The
infinite system can now be divided in groups of layers ("slices") that
only interact with the nearest neighbor groups. The Hamiltonian
matrix elements (and overlaps) for orbitals in the surface slice, and
its interaction with orbitals in underlying layers, are obtained from
the slab calculation. The interactions in the inner slices are taken
from a calculation of bulk Cu. A common energy reference is set by
aligning the Fermi levels of both calculations. We now use the
recursive relation
\begin{equation}
\label{recursive}
G^{ij}(\omega,{\bf k}_\|)[H_{jk}({\bf k}_\|)
-\omega S_{jk}({\bf k}_\|)]=\delta^i_k
\end{equation}
to obtain the Green function in the surface region. We use a method
similar to that described in Ref. \cite{artacho} to iteratively solve
this equation. The matrix
elements are defined such that $G({\bf r},{\bf r^\prime};\omega,{\bf
k}_\|)$= $G^{ij}(\omega,{\bf k}_\|)$ $\phi_i({\bf r})\phi_j({\bf
r^\prime})$, where the $\phi_i$ are the atomic orbitals and the sum
over repeated indices is assumed. Our strategy is then to project the
Green function onto a wave packet $\Psi_R$ localized in the surface
region. The function Im\{$<\Psi_R|\hat{G}(\omega)|\Psi_R>$\} exhibits
well defined peaks at energies $\epsilon_R$ corresponding to the
QWS of the overlayer. The shape and position of these peaks are
independent of the election of $\Psi_R$, as far as $\Psi_R$ has
predominant contributions from the orbitals in the adsorbates. The
peaks have a Lorentzian shape with a width and position that can
be easily extracted using a least square fit~\cite{fit}. Carrying out
this procedure for the two systems studied here we find values of the
elastic width at $\bar{\Gamma}$ equal to 9.4~meV and 7.4~meV for Cs
and Na systems, respectively.

\par For both systems of interest we estimate the inelastic
electron-electron ($e - e$) contribution ($\Gamma_{e-e}$) to the QWS
linewidth by using a model in which the one-electron potential is
constant in a plane parallel to the surface and varies only in the
direction $z$ perpendicular to the surface. We use here the potential
proposed in Ref. \cite{chulkov} for thin films adsorbed on
a Cu(111) substrate with parameters fitted to reproduce the binding
energy of the QWS in the Cs (42 meV) and Na (408 meV) adlayer on
Cu(111). $\Gamma_{e-e}$ is calculated within the GW approximation
\cite{echenique} by employing eigenfunctions and eigenenergies
obtained with this model potential. The computed $\Gamma_{e-e}$
values of less than 0.1 meV and 0.4 meV for a QWS in the Cs
and Na systems, respectively,
demonstrate that inelastic $e - e$ scattering provides only a small
contribution to the QWS linewidth. This latter fact reflects the
difference between electrons and holes: for electrons at the
bottom of the QWS the intraband channel is completely absent, while
it is most important for holes \cite{kliewer}.

Another inelastic contribution, the electron-phonon ($e - ph$) one,
is found to be important for the studied systems. While there is
no detailed information on phonon dispersion of these systems, one
can refer to helium atom scattering measurements for a monolayer of
Cs on Cu(100) \cite{witte}. Assuming that this phonon structure is
also valid for Cs on Cu(111) we evaluate $\Gamma_{e-ph}$ by using
the Einstein model \cite{grimvall} to treat a flat phonon mode with
an energy of 7.5 meV around the $\overline{\Gamma}$ point and the Debye
model \cite{grimvall} for the two acoustic modes (Rayleigh and
longitudinal ones) with maximum energy of 6 meV\@. The $e - ph$
coupling parameter ${\lambda}$ = 0.18 is taken from Ref.
\cite{grimvall} as the maximum bulk value for ${\lambda}$. With
this we estimate $\Gamma_{e-ph}$ = 7.5 meV\@. For Na
on Cu we also assume that the measured maximum phonon frequency for
0.44 ML of Na on Cu(100) can be applied to \ptwo\ Na/Cu(111).
Following Ref. \cite{chulkov} we use
${\lambda}$ = 0.24 and Debye energy $\hbar \omega_{D}$ = 18 meV
yielding $\Gamma_{e-ph}$ = 9.0 meV.

\begin{table}
\begin{center}
\begin{tabular}{c||c|c|c|c}
& $\Gamma_{elastic}$ & $\Gamma_{e-e}$ & $\Gamma_{e-ph}$ &
$\Gamma_{total}$ \\
\hline
\hline
Cs & 9.4 & $<$0.1 & 7.5 & 17 \\
\hline
Na & 7.4 & 0.4 & 9 & 16.8 \\
\end{tabular}
\caption{Line width contributions (in meV) of various scattering channels.
$\Gamma_{elastic}$ elastic,  $\Gamma_{e-e}$ electron-electron,
and $\Gamma_{e-ph}$ electron-phonon scattering. $\Gamma_{total}$
is the resulting total width.}
\label{tab1}
\end{center}
\end{table}

An overview of the different contributions to the line width of
the QWS at the $\overline{\Gamma}$ point is presented in Table
\ref{tab1}. The elastic contribution to the energy width accounts
for approximately 50 \% of the total width. The inelastic
$e - ph$ contribution is comparable. However, the inelastic
$e - e$ contribution is negligible.

\par In summary, our combined experimental and theoretical study shows
that
the dynamics of electrons at ordered overlayers is determined by
the elastic width that appears due to the closing of gaps after back
folding the bands and by inelastic e-ph scattering. However, inelastic e-e
scattering does not play a role.  The relative importance of
the elastic contribution for other ordered superstructures
will depend on details like the density of
states at given points of the substrate SBZ.
Any change of periodicity on going from the
surface to the overlayer structure will lead to a modification of the
scattering channels that determine the dynamics of electrons at the
interface.

\par The Kiel authors thank the Deutsche Forschungsgemeinschaft.
The San Sebastian authors thank Eusko Jaurlaritza,
Euskal Herriko Unibertsitatea, the spanish M.E.C. and the EC grant 
NANOQUANTA for financial support.


\begin{thebibliography}{99}
\bibitem{basis} We have used a double-$\zeta$ polarized basis sets of
pseudo-atomic orbitals. The radii of the orbitals for Cu  were
fixed with an energy shift \cite{SIESTA2} of 200~meV. The basis
set of Cs and
Na corresponds to a +1 ionic configuration, and the orbital radii
were fixed respectively to 9 and 8 Bohrs. For Cs the 5$p$ electrons
were included explicitly in the calculations.
\bibitem{fit} We perform a fit to
$[(\omega-\epsilon_R)^2+(\Gamma_R/2)^2]^{-1}$ in the energy range of
interest.  In order to obtain convergence of the iterative
solution of Eq.~\ref{recursive} it is necessary to evaluate the Green
function outside the real axis. To this end, a small imaginary
part is added to the energy $\omega+i\delta$. To estimate the
physical (elastic) width $\Gamma_{elastic}$ of the QWS, this
artificial width is  substracted from the fitted value $\Gamma_R$. We
have thus $\Gamma_{elastic}$=$\Gamma_R-2\delta$. We have used three
different values of $\delta$, (1.0, 0.5 and 0.25~meV), obtaining
identical values for $\Gamma_{elastic}$ within $\approx 0.1$~meV.
\end{thebibliography}
\end{document}